\documentclass[10pt,aps,twocolumn,showpacs,superscriptaddress]{revtex4}

\usepackage{amsfonts}
\usepackage{graphicx}
\usepackage{amsmath}
\usepackage{amsthm}
\usepackage[colorlinks=true, citecolor=blue, urlcolor=blue ]{hyperref}
\usepackage{graphicx}
\usepackage{textcomp}
\begin{document}
\title{Geometry of quantum state space and quantum correlations}

\author{Prasenjit Deb}
\email{devprasen@gmail.com}
\affiliation{Department of Physics and Center for Astroparticle Physics and Space Science, Bose Institute, Bidhan Nagar
Kolkata - 700091, India.}

\begin{abstract}
Quantum state space is endowed with a metric structure and Riemannian monotone metric is an important geometric entity
defined on such a metric space. Riemannian monotone metrics are 
very useful for information-theoretic and statistical considerations on the quantum state space. 
In this article, considering the quantum state space being spanned by 2x2 density matrices, we determine a particular
Riemannian metric for a state $\rho$ and show that if $\rho$ gets entangled with another quantum state, the 
negativity of the generated entangled state is, upto a constant factor, equals to square root of that particular Riemannian metric 
. Our result clearly relates a geometric quantity to a measure of entanglement. Moreover, the result establishes the
possibility of understanding quantum correlations through geometric approach.
\end{abstract}
\maketitle

\section{Introduction}
Geometric tools are often used to treat physical problems. Undoubtedly, these tools have provided advantage to find out less
trivial and robust physical constraints on physical systems. \emph{Differential geometry} is one such mathematical tool which 
finds a lot of applications in various disciplines. Information theory is among those disciplines where the techniques of
differential geometry have been applied. As a result of which a new discipline, called \emph{Information Geometry} 
emerged  and it got maturity through the
works of Amari,Nagaoka and other mathematicians in the 1980s\cite{Amari}. In their works they applied the methods of 
differential geometry to the field of probability theory, which alongwith statistics is the mathematical base of 
information theory. Initially,the goal of Information geometry was to understand the interplay between the information-theoretic 
quantities and the geometry of probability space by constructing a Riemannian space corresponding to probability
space. Quite obviously it was of fundamental importance to study the Riemannian metrics defined on the space of probability 
distributions.Later, motivated by information geometry, Morozova and \u{C}encov \cite{Cencov} initiated the study of Riemannian monotone
metrics on the state space(Hilbert space) of quantum systems which has been gradually progressed through the works 
of Petz and other authors \cite{Pet1,Pet2,Pet3,Pet4,Isola1,Isola2}. The monotone Riemannian metric corresponding to
Wigner-Yanase-Dyson skew-information \cite{Wigner} was found out in \cite{Pet4} which vividly expresses the relation between geometry of 
space and an information-theoretic quantity of great importance. And not only limited to information-theoretic entities, 
geometric distances(metrics) are also closely related to the quantum  state discrimination problem \cite{Wootters,Caves}.
In\cite{Pinto} the authors have demonstrated that a lower bound for quantum coherence measure can be found out using Riemannian
monotone metric.

\paragraph*{}
On the other hand, quantum correlation is the physical quantity responsible for the non-classical phenomenon exhibited 
by composite quantum systems. Though there are different aspects of quantum correlations, entanglement and discord are 
the two aspects which have been extensively studied due to their immense importance in quantum information processing tasks.
However, till date, quantum correlation is not fully understood. So, the study of quantum correlations demands importance
in quantum information theory. Here we consider entanglement because of the fact that all the measures of entanglement
are monotonic in nature.
\paragraph*{}
Quantum entanglement\cite{Schrodinger,Werner,Horodecki} is one of the bizzare phenomena exhibited by composite quantum 
systems and a resource for quantum information processing tasks, such as teleportation \cite{Tele}, 
dense coding \cite{Dense}, quantum cryptography \cite{Crypto}, state merging\cite{SM}, quantum computation and many more.  
 A composite quantum system $\rho_{AB}$ consisting of subsystems $A$ and $B$ is said 
to be entangled if it can not be written as $\rho_{AB}= \sum_i~p_i \rho_A^i\otimes\rho_B^i$, where $p_i$ are probabilities,
$\rho_A$ and $\rho_B$ are respectively the desity matrices of subsystem $A$ and $B$. If the subsystems are two-level
quantum states then these are termed as \emph{qubits} \cite{Nielsen} in analogy with classical bits and qubits are the
fundamental units in quantum information theory. 
\paragraph*{}
In this article we ask the question; \emph{Is there any connection between the geometry of quantum state space and 
entanglement}? To find the answer of the question we consider the negativity($\mathcal{N}$) \cite{Sanpera,Vidal}measure
which is also connected to some other measures \cite{Toth}. 
Without loss of generality, we consider a generic two-level quantum state $\rho_S$(qubit). 
Quite obviously the dimension of the Hilbert space associated to such a 
state will be 2. It is known that entanglement can be generated between the qubit and another ancillary qubit  by applying 
a global unitary interaction. We consider a particular global unitary and find the generated entangled state.
The amount of entanglement present in the composite state is calculated through negativity measure. We also determine a particular
Riemannian metric for the state $\rho_S$ using a theorem proposed by  Morozova and \u{C}encov.
Interestingly, we find that there is an explicit relation between the negativity of the generated entangled state and
the particular monotone Riemannian metric on the state space of qubits. More precisely to say, the negativity of the
two qubit entangled state is, upto a constant factor, equals to the square root of the monotone metric. 
The result clearly demonstrates the fact that monotone Riemannian metrics on quantum state space are not only connected
to information-theoretic quantities but also to entanglement. Alternatively, we can say that the geometry of quantum state
space finds connection with a non-classical property of the quantum states

\paragraph*{} The rest of the article is arranged as follows. In Section(\ref{sec2}) we provide an overview on Riemannian
metric and Riemannian metrics on matrix space(quantum state space). 
Section(\ref{sec3}) contains a brief description about entanglement generation 
in qubit scenario.
Section(\ref{sec4}) is dedicated to show our results and finally in Section(\ref{sec5}) we conclude our work with discussions.

\section{Riemannian metric and Monotone Riemannian Metrics on matrix space}\label{sec2}
Riemannian space($M$,$g$) is a differentiable manifold(topological space) $M$ embedded with an inner product $g_x$ on the tangent
space $T_xM$ at each point $x$ and $g_x$ varies smoothly from point to point. More precisely if $X$ and $Y$ are two vectors 
on the tangent space passing through $x$ the $x\mapsto g_x\{X(x),Y(x)\}$ is a smooth function. Riemannian metric on a manifold
$M$ is the family of $g_x$.
\paragraph*{}
Morozova and \u{C}encov initiated the study of monotone Riemannian metrices on the space of matrices with the motivation to extend
the geometric approach to quantum setting. They proposed the problem to find monotone Riemannian metrices on the 
quantum state space which is endowed with a metric structure.
The quantum state space is a complex matrix space $\mathcal{H}_n$ of dimension $n$, usually termed as 
\emph{Hilbert space}. The operators in such a space is designated by $n$ x $n$ complex, Hermitian and self-adjoint matrices.
Whereas, the quantum states are defined by positive definite $n$ x $n$ matrices with trace 1, also termed as density matrices.
Hilbert space is an inner product space and the simplest inner product is certainly the Hilbert-Schmidt one, defined as  
\begin{equation}
 \langle X,Y \rangle = \mbox{Tr} (X^*Y)
\end{equation}
where, Tr is the usual matrix trace and $X,Y\in \mathcal{H}_n$. The inner-product in Hilbert space is indeed unitarily invariant, i.e.
$\langle X,Y \rangle = \langle UXU^\dagger, UYU^\dagger \rangle$ and this property is so strong that it determines the 
Hilbert-Schmidt inner-product upto a constant multiple.
\paragraph*{}
Now, by making the inner-products depending on quantum states($\rho$),  Riemannian metrics can be determined on the quantum state space
in the following way. Let for every $A,B\in \mathcal{H}_n$, for every $\rho\in\mathcal{M}_n$, and for every $n\in N $, a complex quantity
$K_{\rho}(A,B)$ is given, where $\mathcal{M}_n$ is the set of all positive definite matrices with trace 1. The complex
quantity $K_{\rho}(A,B)$ will be a metric if the following conditions hold \cite{Pet1}:
\begin{enumerate}
 \item[(a)] $(A,B)\mapsto K_{\rho}(A,B)$ is sesquilinear.
 \item[(b)] $ K_{\rho}(A,A)\geq 0$, and the equality holds \emph{iff} $A=0$
 \item[(c)] $\rho\mapsto K_{\rho}(A,A)$ is continuous on $\mathcal{M}_n$ for every $A$
\end{enumerate}

The family of the metrics $K_{\rho}(A,B)$ with the above mentioned properties constitute a Riemannian metric on the 
differentiable manifold formed by the density matrices. The Riemannian metric will be monotone if 
\begin{enumerate}
 \item [(d)]Under completely positive trace preserving(CPTP) map
$K_{\rho}(A,A)$ is contractive, i.e. $K_{\Lambda(\rho)}(A,A)\leq K_{\rho}(A,A)$ for every $\Lambda$, $\rho$ and $A$; 
$\Lambda(.)$ being the CPTP map.
\end{enumerate}
For clear illustration of the metric $K_{\rho}(A,B)$, it is important to focus on the geometry of the quantum state space. $K_{\rho}(A,B)$
is basically the inner product on the tangent space $T_{\rho}$ and $A$,$B$ are the two tangent vectors. Considering the Hilbert
space $\mathcal{H}_n$ to be finite dimensional, let us denote the set of all Hertmitian operators on $\mathcal{H}_n$ by 
\begin{equation}
 \mathcal{A}=\{A\lvert A=A^*\}
\end{equation}
and according to definition of $\mathcal{M}_n$;
\begin{equation}
 \mathcal{M}_n= \{\rho \lvert \rho=\rho^* \geq 0~ \mbox{and} ~\mbox{Tr} \rho=1\}
\end{equation}
The tangent space $T_{\rho}(\mathcal{M}_n)$ of each point $\rho$ may then be identified with
\begin{equation}
 \mathcal{A}_0= \{A\lvert A \in \mathcal{A} ~ \mbox{and}~ \mbox{Tr}A=0\}
\end{equation}
It can be shown that if $\mathcal{K}$ is an operator and $\mathcal{K}\in \mathcal{A}$, then $i [\rho, \mathcal{K}]$ will be an 
ordinary element of the tangent space, that is, $i [\rho, \mathcal{K}]\in\mathcal{A}_0$\cite{Amari}. So, by identifying tangent vectors
Riemannian metric can be defined on the differential manifold formed by the density matrices and upon satisfying the condition
(d) the metric will be called monotone Riemannian metric.

\paragraph*{}
Though Morozova and \u{C}encov were unable to find any monotone Riemannian metric, they provided a useful theorem. Later, Petz and other
authors were  able to find monotone metrics by introducing operator montone functions and their works showed that there is
an abundance of montone metrics on the space of self-adjoint matrices \cite{Pet2,Pet3}. For our purpose we will make use of the theorem
provided by Morozova and \u{C}encov, which can be stated as,
\\\\\
THEOREM\cite{Cencov,Pet1}:\emph{Assume that for every $D\in\mathcal{M}_n$ a real bilinear form $K_D^\prime$ is given on
the n-by-n self-adjoint matrices such that the conditions} (b),(c) and (d) \emph{are satisfied for self-adjoint A. 
Then there exists a positive continuous
function c($\lambda$,$\mu$) and a constant C with the following property: If D is diagonal with respect to the matrix units
$E_{ij}$, i.e. $ D = \sum_i \lambda_i E_{ii}$, then}

\begin{equation}
 K^\prime(A,A)= C\sum_{i=1}^n\lambda_i^{-1}A_{ii}^2 + 2\sum_{i<j} \lvert A_{ij} \lvert ^2 c(\lambda_i, \lambda_j).\label{metric}
\end{equation}
\emph{for every self-adjoint $A=(A_{ij})$. Moreover if c is symmetric in its two variables, $c(\lambda,\lambda)= C \lambda^{-1}$
 and $c(t\lambda,t\mu)= t^{-1}c(\lambda,\mu)$ } 
\\\\\
The function $c(\lambda,\mu)$ is generally termed as \emph{Morozova-\u{C}encov} function.
It can be concluded from the above stated theorem that when $\mathcal{M}_n$ is considered as a differentiable manifold,
the Riemannian metric must be a linear bilinear form and the tangent vectors may be indentified with self-adjoint matrices.
Also $K^\prime(A,A)$ for all $D$ and all self-adjoint $A$ can be derived from the theorem.

\section{Entanglement generation in qubit scenario}\label{sec3}
Let us consider a generic qubit $\rho_S$, generally expressed as:
\begin{equation}
 \rho_S=\frac{1}{2}(\mathbf{1}_2+\vec{n}.\vec{\sigma})
\end{equation} where, $\vec{n}\equiv(n_x,n_y,n_z)$ is a vector in $\mathbb{R}^3$ 
with $|\vec{n}|^2\le 1$ and $\vec{\sigma}:=(\sigma_x,\sigma_y,\sigma_z)$ 
with $\sigma_i|i=x,y,z$ being the Pauli matrices. Taking another ancillary qubit, say $\lvert 0\rangle_A$, which is an 
eigen state of $\sigma_z$, a unitary interaction is switched on over the product state $\rho_S\otimes\lvert 0\rangle_A$.
The global unitary $U_{SA}$ acts on the initial product state as:
\begin{eqnarray}\label{int}
U_{SA}(|0\rangle_S\otimes|0\rangle_A)=|0\rangle_S\otimes|0\rangle_A\nonumber\\
U_{SA}(|1\rangle_S\otimes|0\rangle_A)=|1\rangle_S\otimes|1\rangle_A.
\end{eqnarray}
So, after the application of the unitary the resulting state becomes
\begin{eqnarray}\label{state}
\rho_{SA}&=& U_{SA}\left(\frac{1}{2}(\mathbf{1}+\vec{n}.\vec{\sigma})_s\otimes|0\rangle_M \langle0|\right) U_{SA}^\dagger\nonumber\\
&=& \frac{1+n_z}{2}|0\rangle_S\langle 0|\otimes|0\rangle_M\langle 0|\nonumber\\
&&+\frac{n_x+in_y}{2}|1\rangle_S\langle 0|\otimes|1\rangle_M\langle 0|\nonumber\\
&&+\frac{n_x-in_y}{2}|0\rangle_S\langle 1|\otimes|0\rangle_M\langle 1|\nonumber\\
&&+\frac{1-n_z}{2}|1\rangle_S\langle  1|\otimes|1\rangle_M\langle 1|. \label{joint-state}
\end{eqnarray}
The unitary operator corresponding to the evolution of the product state $\rho_S\otimes\lvert 0\rangle_A$ can be
expressed in terms of the total Hamiltonian
($H_{tot}$) as $U(t):=\exp^{(-i\frac{H_{tot}t}{\hbar})}$ and 
$H_{tot}=H_S\otimes\mathbf{1}_M+\mathbf{1}_S\otimes H_M+H_{int}$. So it is clear that the initial composite product state will 
retain its product form
if $H_{int}=0$.
\paragraph*{}
We are interested to find out the amount of entanglement in the state $\rho_{SA}$ and among different measures of entanglement
we consider the negativity measure for our purpose. Negativity is given by \cite{Sanpera,Vidal}:
\begin{equation}
\mathcal{N}(\rho_{AB})=\frac{||\rho^{T_A}_{AB}||_1-1}{2}.\label{neg-formula}
\end{equation}
where $T_A$ denotes partial transpose with respect to the subsystem $A$, $\lambda_i$'s denote the eigenvalues 
of $\rho^{T_A}_{AB}$ and $||X||_1=\mbox{Tr}|X|=\mbox{Tr}\sqrt{X^{\dagger}X}$ be the trace-norm of an operator. 
Using Eqn(\ref{neg-formula})the negativity of the state in Eqn(\ref{joint-state}) is found to be
\begin{equation}
 \mathcal{N}(\rho_{SA})=\left(1-\sqrt{2\mathcal{M}(\rho_S)}\right)^{-\frac{1}{2}}
\frac{(1-\sqrt{1-|\vec{n}|^2})^\frac{1}{2}(n^2_x+n^2_y)^{\frac{1}{2}}}{2} \label{neg-state}
\end{equation}
for $|\vec{n}|\neq 0$ and
\begin{equation}
 \mathcal{N}(\rho_{SA})=0,~~~~ \mbox{if},~ \lvert n \lvert =0.\label{nil-neg}
\end{equation}
where, $\mathcal{M}(\rho_S)$ is the mixedness of the state $\rho_S$ and is given by:
\begin{eqnarray}
 \mathcal{M}(\rho_S^{in})&=&\mbox{Tr}(\rho_S^{in})-\mbox{Tr}(\rho_S^{in})^2\nonumber\\
&=&\frac{1}{2}(1-|\vec{n}|^2).
\end{eqnarray}

From Eqn(\ref{nil-neg}) it is clear that negativity of the state will be zero if $\rho_S$ is maximally mixed, i.e. 
$\rho=\frac{\mathbf{1}_2}{2}$.

\section{Results}\label{sec4}
Now we show that the negativity $\mathcal{N}$ of the state $\rho_{SA}$ is, upto a constant, equal to a Riemannian metric on 
the quantum state space of dimension 2. Let us consider the self-adjoint, Hermitian operator $\sigma_z$. 
Then the operator $i[\rho_S,\sigma_z]$ is also a Hermitian, self-adjoint operator and it is an element of the tangent space of
density matrices differentiable manifold \cite{Isola2}. Our aim is to find the metric $K_{\rho_S}(A,B)$, where $A=B= i[\rho_S,\sigma_z]$.
\paragraph*{}
The state $\rho_S$ can be represented in the diagonal form as:
\begin{eqnarray}
 \rho_S&=& \frac{1}{2}(\mathbf{1}+|\vec{n}|\hat{n}.\vec{\sigma})\nonumber\\
&=&\frac{1+|\vec{n}|}{2}\frac{1}{2}(\mathbf{1}+\hat{n}.\vec{\sigma})+\frac{1-|\vec{n}|}{2}\frac{1}{2}
(\mathbf{1}-\hat{n}.\vec{\sigma}).\label{qubit-diag}
\end{eqnarray}
where, $\hat n= \frac{\vec n}{\lvert\vec n\lvert}$.~
Moreover, the matrix representation of self-adjoint operator $i[\rho_S,\sigma_z]$ will be:
\begin{equation}
 i[\rho_S,\sigma_z]=\lvert\vec n\lvert \begin{pmatrix} 0 &(-n_y-in_x) \\ (in_x-n_y)& 0 \end{pmatrix}.\label{matrix-rep}
\end{equation}
It is easy to verify that for the operator represented in Eqn(\ref{matrix-rep}) the first summation term in Eqn(\ref{metric})
will be zero, i.e,
\begin{equation}
 C\sum_{i=1}^2\lambda_i^{-1}A_{ii}^2=0
\end{equation}
where, $A_{ii}$ represent the diagonal elements of the matrix corresponding to the operator $i[\rho_S,\sigma_z]$. Therefor,
\begin{equation}
 K_{\rho_S}(i[\rho_S,\sigma_z],i[\rho_S,\sigma_z])= 2\sum_{i<j} \lvert A_{ij} \lvert ^2 c(\lambda_i, \lambda_j)
\end{equation}
$A_{ij}$ being the off-diagonal elements of the matrix given in Eqn(\ref{matrix-rep}) and $c(\lambda_i,\lambda_j)$ is 
the Morozova-\u{C}encov
function \cite{Cencov,Pet1}. For our purpose we consider one of the functions proposed originally by Morozova and \u{C}encov \cite{Cencov}; 
we take
\begin{equation}
 c(\lambda_i,\lambda_j)=\left(\frac{2}{\sqrt{\lambda_i}+\sqrt{\lambda_j}}\right)^2.\label{mc-function}
\end{equation}
Therefore, using Eqns[\ref{qubit-diag},\ref{matrix-rep}] and Eqn(\ref{mc-function}), we get
\begin{eqnarray}
 K_{\rho_S}(i[\rho_S,\sigma_z],i[\rho_S,\sigma_z])&=& 2.\lvert A_{12}\lvert^2
 \left(\frac{2}{\sqrt{\frac{1+\lvert\vec{n}\lvert}{2}}+\sqrt{\frac{1-\lvert\vec{n}\lvert}{2}}}\right)^2\nonumber\\
 &=& 32\lvert A_{12}\lvert^2\frac{(1-\sqrt{1-\lvert\vec{n}\lvert^2)}}{\lvert\vec{n}\lvert^2}\nonumber\\
 &=& 32(n_x^2+n_y^2)(1-\sqrt{1-\lvert\vec{n}\lvert^2)}\label{metric1}
\end{eqnarray}
It is to be noted that we have considered the general density matrix $\rho_S$ and determined the Riemannian metric for 
diag[$\lambda_1,\lambda_2$]; where $\lambda_1$ and $\lambda_2$ are the eigenvalues of $\rho_S$. Nevertheless, the evaluated
metric is the one which we want and the unitary covariance $K_{\rho_S}(A,A)=K_{U^*\rho_SU}(U^*AU,U^*AU)$ confirms the fact \cite{Pet1}. 
\\
\paragraph*{}
Finally, using Eqn(\ref{neg-state}) and Eqn(\ref{metric1}) we get;
\begin{equation}
 \mathcal{N}(\rho_{SA})= \mbox{A}\sqrt{K_{\rho_S}(i[\rho_S,\sigma_z],i[\rho_S,\sigma_z])}.\label{neg-metric}
\end{equation}
\\
where, A~=~$2\sqrt{2}\left(1-\sqrt{2\mathcal{M}(\rho_S)}\right)^{-\frac{1}{2}}$
\\\\
The above equation is the main result of this article. It is worth noticing that the above equation vividly depicts a 
relation between a geometric entity of the differential manifold formed by density matrices and a measure of entanglement.
Now, as the above equation is a manifestation of equality relation between two different entities, it is necessary to check
the validity of the equality and this can be verified with the following arguments;
\begin{itemize}
 \item Negativity($\mathcal{N}$) is unitarily invariant, i.e. 
 $\mathcal{N}(\rho_{SA})=\mathcal{N}(U_{SA}\rho_{SA}U^\dagger_{SA})$ and it is also a monotonic function;  
 $\mathcal{N}(\rho_{SA})\geq \mathcal{N}(\Lambda(\rho_{SA}))$, where $\Lambda$ is a CPTP map.
 \item Riemannian metric $K_{\rho_S}(i[\rho_S,\sigma_z],i[\rho_S,\sigma_z])$ is a monotone metric due to fulfilment of conditions
 (d) provided in Sec[\ref{sec2}] and monotonocity  includes the unitary covariance of the metric.
\end{itemize}
From the two arguments it can be concluded undoubtedly that our result provides an elegant relation between the geometry 
of the quantum state space and a non-classical phenomena exhibited by composite quantum system.
\paragraph*{}
It is important to highlight that Eqn(\ref{neg-metric}) encompasses some important facts directly. The Riemannian metric
$K_{\rho}(A,A)=0$ \emph{iff} $A=0$. So, $i[\rho_S,\sigma_z]=0$ implies $K_{\rho_S}(i[\rho_S,\sigma_z],i[\rho_S,\sigma_z])=0$,
which in turn gives $\mathcal{N}_{SA}=0$. All these are in agreement with the physical fact that if the state $\rho_S$ is
a mixture of the eigen states of the observable $\sigma_z$ then no entanglement can be generated
between the state and the ancilla by applying $\sigma_z$ interaction. The metric that we have calculated was identified as 
Wigner-Yanase(WY) skew information and WY skew-information is a good measure of coherence\cite{Braum,Giro}. Hence, our result identifies a 
relation between entanglement and coherence, thereby reconfirming the fact that in order to generate entanglement between two
qubits, at least one qubit must be in coherent state\cite{Alex}.
\paragraph*{}
Though we have considered a special global unitary $U_{SA}$ in the entanglement generation scenario, as
well as two specific tangent vectors(self-adjoint operators), i.e. $A=B= i[\rho_S,\sigma_z]$ to determine the monotone
Riemannian metric on the quantum state space, nevertheless, the relation given in Eqn(\ref{neg-metric}) is not restricted
to be a special case. Consider $A=B= i[\rho_S,\sigma_x]$ as the tangent elements and the ancillary qubit to be
$\frac{1}{\sqrt{2}}(\lvert + \rangle + \lvert -\rangle)_A$.
Further assume that the global unitary
acts as :
\begin{eqnarray}\label{int}
U_{SA}(|+\rangle_S\otimes|+\rangle_A)=|+\rangle_S\otimes|+\rangle_A\nonumber\\
U_{SA}(|+\rangle_S\otimes|-\rangle_A)=|-\rangle_S\otimes|-\rangle_A.
\end{eqnarray}
Now, if the Riemannian monotone metric is determined and negativity of the generated entangled state is calculated then we
can have a similar relation between these two quantities just like in Eqn(\ref{neg-metric}), i.e.
\begin{equation}
 \mathcal{N}(\rho_{SA})= \mbox{A}\sqrt{K_{\rho_S}(i[\rho_S,\sigma_x],i[\rho_S,\sigma_x])}
\end{equation}

\section{Conclusions}\label{sec5}
Quantum state space is endowed with a metric structure and monotone Riemannian metrics are important candidates for 
quantum-information-theorectic considerations on such a space. In this article we consider 2-dimensional Hilbert space
and show that a particular Riemannian metric is, upto a constant, equals to a measure of entanglement. To express more
precisely, if a general qubit is considered and entanglement is generated between the qubit and an ancilla by applying
interaction, then the negativity of entanglement is, upto a constant, equals to a Riemannian metric. The metric that we
have considered was shown to be the Wigner-Yanase skew-information. Moreover, Wigner-Yanase skew-information is a good measure
of coherence. So, our result also establishes a connection between coherence and entanglement. It is important to emphasize
that so far Riemannian metrics on quantum state space found relation with information-theoretic quantities only, whereas, our
result vividly shows that a monotone Riemannian metric in a manifold of dimension 2 can also be related to negativity of
entanglement, which is also a monotone function. The result established in this article positively illustrates the fact 
that quantum correlations can be studied using geometric approach. 

\paragraph*{}
~~~~~~~~~~~~~~~~~~{\bf Acknowledgement}
\\
The author would like to acknowledge DST, Govt of India, for the financial support.

\end{document}